\documentclass[11pt, a4paper]{article}
\usepackage[a4paper,left=2cm,right=2cm]{geometry}
\usepackage{amsfonts} 	
\usepackage{amsmath}
\usepackage{derivative}
\usepackage{color}
\usepackage[sort&compress, numbers]{natbib}
\usepackage{graphicx}
\usepackage{authblk}

\newcommand{\wand}{\quad\text{and}\quad}
\newcommand{\wfor}{\quad\text{for}\quad}

\newcommand{\was}{\quad\text{as}\quad}
\newcommand{\wat}{\quad\text{at}\quad}
\newcommand{\LR}[1]{\left(#1\right)}
\newcommand{\tr}[1]{\text{tr}\left(#1\right)}
\newcommand{\LRB}[1]{\left[#1\right]}

\newcommand{\LRBR}[1]{\left\{#1\right\}}

\newcommand{\p}{\partial}

\newcommand{\eoe}[2]{\mathbf{e}_{#1}\otimes\mathbf{e}_{#2}}
\newcommand{\XM}{\mathcal{X}_1}
\newcommand{\Bi}{\text{Bi}}
\renewcommand{\i}{\text{i}}
\newcommand{\B}[1]{\overline{#1}}
\newcommand{\W}[1]{\widetilde{#1}}

\newcommand{\nx}{\nabla_{\mathbf{x}}}

\numberwithin{equation}{section}

\title{A nonlinear beam model for photoresponsive thermoelastic
solids driven by localised heating}

\author[1]{William T.~Simpkins\thanks{will.simpkins@bristol.ac.uk}}
\author[2]{Matteo Taffetani\thanks{matteo.taffetani@ed.ac.uk}}
\author[1]{Matthew~G.~Hennessy\thanks{matthew.hennessy@bristol.ac.uk}}

\affil[1]{School of Engineering Mathematics and Technology, University of Bristol, Bristol, BS8 1TW, UK}
\affil[2]{School of Engineering, University of Edinburgh, Edinburgh, EH9 3FB, UK}
\date{}

\begin{document}

\maketitle

\begin{abstract}
    Asymptotic methods are used to derive a 
    geometrically nonlinear beam model for
    thermoelastic solids with a spatially localised
    heat source.  The asymptotic reduction is based on 
    collapsing the heated region to a point.  Away from the
    point of heating, the governing equations reduce to 
    a pair of
    beam equations with nonlinear
    von K\'arm\'an strains.  The effects of the localised heat
    source 
    are captured through asymptotically consistent
    jump conditions that hold at the point of heating.  The 
    model accounts for changes in beam length due to longitudinal thermal expansion and bending moments produced by transverse thermal
    gradients. The model is used to study light-induced
    actuation of photoresponsive hydrogel beams with
    localised heating arising from laser irradiation.  Two loading scenarios are considered.
    In the first, the ends of the beam are assumed to be free,
    resulting in a V-shaped deformation upon heating.  An
    analytical expression for the fold angle of the V is
    provided.  In the second, the beam is assumed to be
    in a pre-buckled configuration due to clamped end
    conditions.  The critical conditions leading to light-driven 
    snap-through are calculated.  Offsetting the laser from 
    the mid-point of the beam is found to inhibit the onset
    of snap through.
    \end{abstract}

\section{Introduction}


Spatially localised thermal stimuli provide a flexible route for actuating slender elastic structures~\cite{Jiang2023,Hippler2019,Song2019}.  Optical heating by a laser is an attractive mode of thermal stimulus because the position, intensity, and duration of irradiation can be varied to produce a controlled material response without contacting the structure.  In these photoresponsive materials, the energy of absorbed light is converted into heat.  The resulting temperature increase can then drive deformation through thermal expansion.  However, for photo-thermo-responsive hydrogels, heating often leads to gel deswelling and thus gives rise to an effective thermal contraction.  In slender structures, weak thermal gradients that are spatially localised can induced global shape changes through bending, folding, buckling, and snap through~\cite{Light,Cao2022}, which can be harnessed in soft robots~\cite{Ni2023}, metamaterials~\cite{Hippler2019,Song2019}, biomedical devices~\cite{Qian2023}, and actuators~\cite{huang2025}.  Despite the wide variety of mechanisms that can drive shape change, programming the response of structures to thermal stimuli remains a key challenge.  

Mathematical modelling can be used to predict the response of materials
to localised thermal stimuli.  However, the complexity of three-dimensional nonlinear thermoelasticity can make numerical simulations computationally expensive and analytical studies intractable.  Thermoelastic beam theories overcome both obstacles by providing dimensionally reduced equations that retain the key physics of the system.  As a result, there is a wide body of literature in which thermoelastic beam models are used to study temperature-driven morphing of slender structures due to longitudinal thermal expansion~\cite{Li2002,Vaz2003}
or bending moments arising from transverse thermal gradients~\cite{Liu2012, Timoshenko1925,Li2006,Li2006a,Paul2016}.
These two effects are seldom considered together.
Other works have examined thermally induced buckling and snap through in beams with prescribed temperature profiles~\cite{Li2007,Smith2025}.  In many physical systems, however, the temperature field is not known \emph{a priori} and must be determined as part of the problem.   

Beam theories for photoresponsive thermoelastic media have recently been developed for liquid crystal elastomers~\cite{Korner2020,Norouzikudiani2023}.  These models extend thermoelasticity by capturing the propagation and absorption of light through the material using the Beer--Lambert law.  If the material deforms into the path of the laser, ray tracing can be used to calculate the resulting shadow and determine if the local rate of heat generation must be reduced accordingly~\cite{Norouzikudiani2024}.  Models for photoresponsive thermoelastic media are usually based on the assumption that the entire structure is being irradiated.  However, recent experiments on slender photo-thermo-responsive hydrogels~\cite{Light} have shown how shape change and snap-through instabilities can be induced by localised irradiation. The localisation of light,
and thus heating, suggests there is scope for simplifying existing thermoelastic beam theories.

In this paper, we derive a nonlinear beam model for a photoresponsive thermoelastic material with a localised heat source.  Unlike
previous works, we derive the beam model directly from the three-dimensional
equations of nonlinear thermoelasticity, accounting for light absorption using
the differential form of the Beer--Lambert law.  By using asymptotic methods to
systematically reduce the governing equations, we derive asymptotically consistent
boundary conditions that simultaneously capture the length change of the beam 
due to longitudinal thermal expansion and bending moments induced by transverse temperature gradients. 

The paper is organised as follows. Sec.~\ref{Problem formulation} presents the full three-dimensional model, which is then non-dimensionalised in Sec.~\ref{Scaling}.  An asymptotic reduction of the model is carried out in Sec.~\ref{Asymptotics}.  The final form of the beam model is presented in Sec.~\ref{Reduced model}.  The model is then used to study morphing of photo-thermal-responsive hydrogels in Sec.~\ref{Static analysis}.  The paper
concludes in Sec.~\ref{Conclusion}.

\section{Governing equations}\label{Problem formulation}

\begin{figure}
    \centering
    \includegraphics[width=0.95\linewidth]{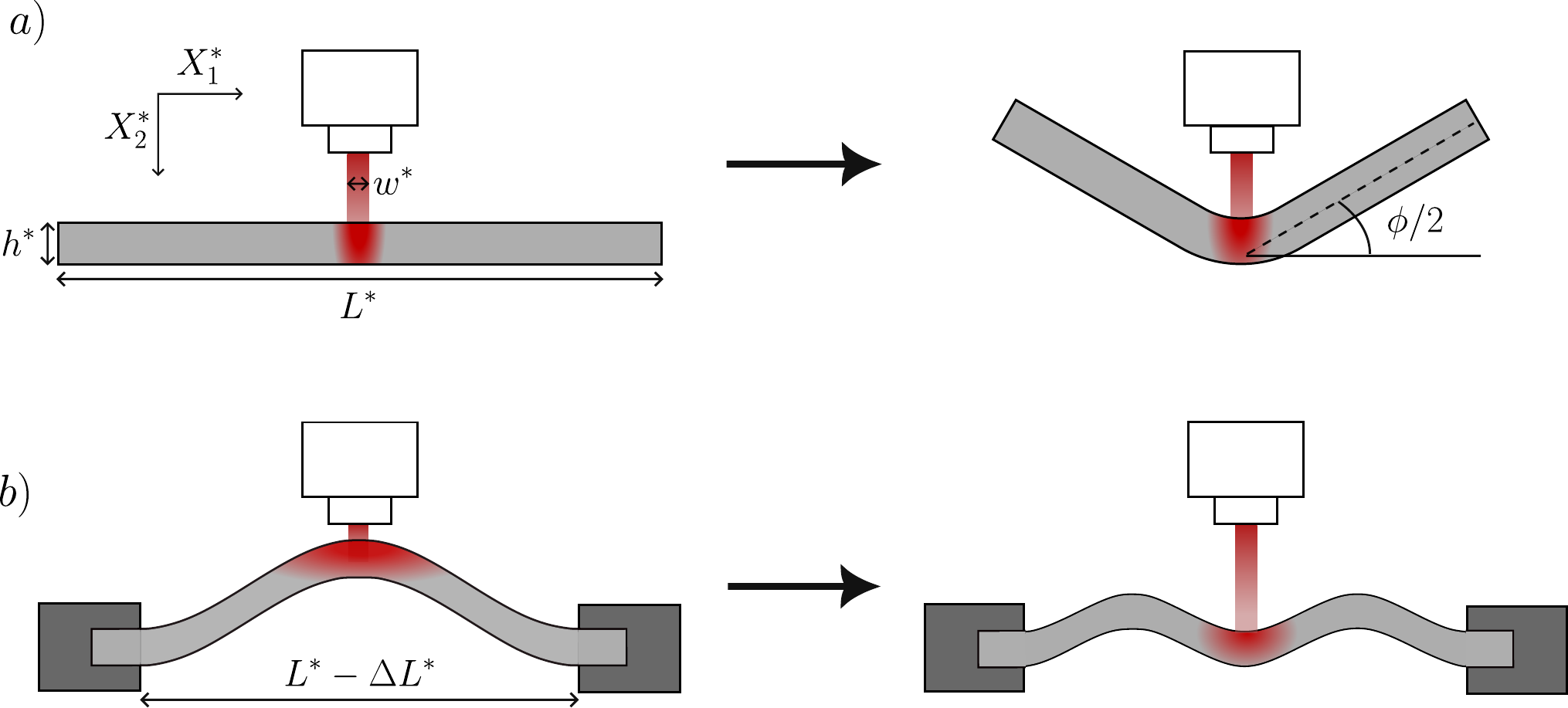}
    \caption{Light-induced deformation of a thermoelastic beam
    with negative thermal expansion coefficient due to laser irradiation with (a) free and (b) clamped ends.  The beam has length $L^*$, height $h^*$, and width $h^*$.  The radius of the laser is $w^*$.
    }
    \label{fig:Schematic}
\end{figure}

We consider a beam of length $L^*$ and a square cross-section of side length $h^*$.  The upper surface of the beam is irradiated by a laser, which provides a localised source of heating.  Due to thermal strains, the beam deforms when irradiated.  The path 
of the laser is assumed to be perpendicular to the undeformed
surface of the beam.  Two loading scenarios are considered based
on experiments involving photo-thermo-responsive hydrogels~\cite{Light}.
In the first, the ends of the beam are free, so it bends into
a V shape when irradiated (Fig.~\ref{fig:Schematic}~(a)).
In the second, the beam is assumed to in a buckled state due
to pre-compression.  Irradiation then triggers a snap-through
instability (Fig.~\ref{fig:Schematic}~(b)).


\subsection{Kinematics and balance laws}

The beam is modelled using the equations of nonlinear
thermoelasticity~\cite{Gurtin2010}.  
The equations are formulated using 
Lagrangian coordinates $\mathbf{X}^*=(X_1^*,X_2^*,X_3^*)$ fixed to the material body.  
This body deforms after some time $t^*$ to the current configuration with
Eulerian coordinates $\mathbf{x}^*=(x_1^*,x_2^*,x_3^*)$ fixed
in the lab frame. 
The deformation is described by the function
\begin{equation}
    \mathbf{x}^*=\chi^*\LR{\mathbf{X}^*,t^*}=\mathbf{X}^*+\mathbf{u}^*\LR{\mathbf{X}^*,t^*},
\end{equation}
where $\mathbf{u}^*=(u_1^*,u_2^*,u_3^*)$ describes the displacement of a material element from $\mathbf{X}^*$ to $\mathbf{x}^*$. Here and throughout, asterisks denote dimensional quantities. The Lagrangian coordinates are defined on the intervals
\begin{equation}
    X^*_1\in \LRB{-\frac{L^*}2,\frac{L^*}2} \wand X_\alpha ^*\in\LRB{-\frac{h^*}2,\frac{h^*}2}\wfor \alpha=2,3.
\end{equation}
We refer to the $X_1$ coordinate, which is aligned with the $\mathbf{E}_1$ axis, as the longitudinal direction and the $X_\alpha$ coordinates, aligned with the $\mathbf{E}_\alpha$ axes, as the transverse directions ($\alpha = 2,\,3$).  In general, Greek indices will refer to both transverse directions interchangeably.
The deformation gradient tensor is
\begin{equation}\label{F definition}
    \mathbf{F}^*= \mathbf{I}+\pdv{\mathbf{u}^*}{\mathbf{X}^*},
\end{equation}
and describes the distortion of material elements.  The
determinant of the deformation gradient tensor,
$J^* = \det \mathbf{F}^*$, describes the volume change 
of material elements.

The governing equations for the beam consist of the balance
of linear momentum
\begin{equation}\label{TE:lin mom dim}
    \rho_0^*\pdv[2]{\mathbf{u}^*}{t^*} = \nabla^*\!\cdot \mathbf{P}^*,
\end{equation}
the balance of angular momentum
\begin{equation}\label{TE:ang mom dim}
    \mathbf{P}^*\LR{\mathbf{F}^*}^T =\mathbf{F}^*\LR{\mathbf{P}^*}^T,
\end{equation}
and the energy balance 
\begin{equation}\label{TE:energy dim}
    \rho_0^*\pdv{e^*}{t^*} = \mathbf{P}^*\!:\!\pdv{\mathbf{F}^*}{t^*}-\nabla^*\!\cdot\!\mathbf{Q}^*+R^*,
\end{equation}
where $\rho_0^*$, $\mathbf{P}^*$, $e^*$, $\mathbf{Q}^*$, and $R^*$ denote the initial density, first Piola--Kirchhoff stress tensor, internal energy per unit mass, heat flux, and heat supply function, respectively.

The thermal distortion of the body is included through a multiplicative decomposition of the deformation gradient tensor
\begin{equation}\label{F decomposition}
    \mathbf{F}^*=\mathbf{F}_\text{e}^*\mathbf{F}_\theta^*,
\end{equation}
where $\mathbf{F}_\theta^*$ describes the thermal distortion tensor and $\mathbf{F}^*_\text{e}$ is the elastic part of the deformation gradient tensor. Multiplicative decompositions of this type are commonly used to represent stress-free thermal distortion in finite-strain elasticity \cite{Lubarda2004}.
The thermal distortion need not be compatible; that is, it does not have to be the derivative of an admissible displacement, which allows the thermal distortion to be prescribed independently of mechanical compatibility. The elastic part $\mathbf{F}_\text{e}^*$ ensures the total (observed) deformation is compatible, allowing the total deformation gradient to be written as in \eqref{F definition}. Equivalently, \eqref{F decomposition} can be written as 
\begin{equation}
    \mathbf{F}_\text{e}^* = \mathbf{F}^*\LR{\mathbf{F}_\theta^*}^{-1},
\end{equation}
where we assume that $\mathbf{F}_\theta^*$ is invertible. We further assume that the elastic response is incompressible, so we have
\begin{equation}\label{incompressibility condition}
    \det\LR{\mathbf{F}_\text{e}^*}=1.
\end{equation}
Thus, volumetric changes of material elements is solely due
to thermal expansion, $J^* = \det \mathbf{F}_\theta^*$.
The thermal distortion is assumed to be orthotropic, leading to \cite{Brunner2024}
\begin{equation}
    \mathbf{F}_\theta^* = \text{diag}\LR{e^{\gamma_1^*(\theta^*-\theta_0^*)},e^{\gamma_2^*(\theta^*-\theta_0^*)},e^{\gamma_3^*(\theta^*-\theta_0^*)}}
\end{equation}
where $\theta^*$ is the material temperature, $\theta_0^*$ is the ambient stress-free temperature, and $\gamma^*_i$ are the coefficients of thermal expansion. This form allows us to describe anisotropic thermal expansion and contraction, which is a feature of some materials~\cite{Light}.

For convenience, we define here the right Cauchy-Green tensors for the total, elastic, and thermal deformation as
\begin{equation}
    \mathbf{C}^* = \LR{\mathbf{F}^*}^T\mathbf{F}^*,\quad \mathbf{C}_\text{e}^* = \LR{\mathbf{F}_\text{e}^*}^T\mathbf{F}_\text{e}^*,\wand \mathbf{C}_\theta^* = \LR{\mathbf{F}_\theta^*}^T\mathbf{F}_\theta^*.
\end{equation}

\subsection{Constitutive assumptions}

To define the constitutive behaviour, we introduce the Helmholtz free energy per unit reference volume 
\begin{equation}
    \Psi^* = \rho_0^*\LR{e^*-\theta^*\eta^*},
\end{equation}
where $\eta^*$ is the entropy per unit mass. We assume that the material response is determined by a free energy of the form
\begin{equation}
    \Psi^*\LR{\mathbf{F}^*,\theta^*,\hat{p}^*} = \Psi^*_\theta(\theta^*)+\Psi_\text{e}^*\LR{\mathbf{F}_\text{e}^*,\theta^*}-p^*\LR{\det(\mathbf{F}_\text{e}^*)-1},
\end{equation}
where $\Psi_\theta^*$ is the caloric contribution to the free energy, $\Psi_\text{e}^*$ is the elastic contribution, and $p^*$ is a Lagrangian multiplier which enforces elastic incompressibility. We take the caloric contribution to be of the form \cite{Brunner2024}
\begin{equation}
    \Psi^*_\theta(\theta^*) = \rho_0^*c_\theta^*\LR{\theta^*-\theta_0^*-\theta^*\ln\LR{\frac{\theta^*}{\theta_0^*}}},
\end{equation}
where $c_\theta^*$ is the specific heat capacity. For the elastic part, we use a compressible neo-Hookean form \cite{Holzapfel2002}
\begin{equation}
    \Psi^*_\text{e}\LR{\mathbf{F}_\text{e}^*,\theta^*} = \frac{E^*
    }{3}\LRB{\tr{\mathbf{C}_\text{e}^*}-3-\ln\LR{\det\LR{\mathbf{C}_\text{e}^*}}},
\end{equation}
where $E^*$ is the modulus, which we assume to be temperature independent due to the expected smallness of temperature
increases.

The constitutive equations are required to satisfy the Clausius-Duhem inequality.  By applying the Coleman-Noll procedure~\cite{Coleman1963,Gurtin2010}, the constitutive
relations are found to be
\begin{subequations}
\begin{align}
    \mathbf{P}^* &= \pdv{\Psi_\text{e}^*}{\mathbf{F}^*}-p^*\LR{\mathbf{F}^*}^{-T},\\
    \mathbf{Q}^* &= -\mathbf{K}^*\nabla^*\theta^*,\\
    \rho_0^*\eta^* &= -\pdv{\Psi_\theta^*}{\theta^*}-\pdv{\Psi_\text{e}^*}{\theta^*}-p^*\LR{\mathbf{F}_\theta^*}^{-T}:\pdv{\mathbf{F}^*_\theta}{\theta^*},\label{e identity}
\end{align}
\end{subequations}
where $\mathbf{K}^* = k^* J^* (\mathbf{C}^*)^{-1}$ is the thermal conductivity tensor, with $k^*$ denoting the isotropic
thermal conductivity in the current configuration, and $e^*$ is the entropy per
unit mass.
For the neo-Hookean energy given above, the first Piola--Kirchhoff stress given by
\begin{equation}\label{P dimensional}
    \mathbf{P}^* = \frac{2E^*}{3}\LRB{\mathbf{F}^*\LR{\mathbf{C}_\theta^*}^{-1}-\LR{\mathbf{F}^*}^{-T}}-p^*\LR{\mathbf{F}^*}^{-T}.
\end{equation}
In addition, the energy balance \eqref{TE:energy dim} 
can be converted into an entropy balance~\cite{Brunner2024}
given by
\begin{equation}
    \rho_0^*\theta^*\pdv{\eta^*}{t^*} = \nabla^*\cdot\LR{\mathbf{K}^*\nabla^*\theta^*}+R^*,
\end{equation}
where the entropy $\eta^*$ is defined in \eqref{e identity}.

\subsection{Light propagation and heating}
\label{sec: Geometry and heat source}

We assume that the beam is irradiated by a laser of
power $P^*$.  The laser is assumed to have a 
Gaussian beam of radius $w^*$ that hits the
the upper surface of the material at 
longitudinal coordinate
$x_1^* = X_0^*$ and transverse coordinate $x_3^* = 0$.  
Thus, the laser intensity at the upper surface of the beam
(i.e.\ the incident laser intensity) is given by
\begin{align}
I_0^* = \frac{P^*}{\pi (w^*)^2} \exp\left(-\frac{(x_1^* - X_0^*)^2 + (x_3^*)^2}{(w^*)^2}\right).
\label{eqn:I_0}
\end{align}
As the laser travels through the material, the absorption of light produces heat.  The rate of heat generation per unit volume is given by
\begin{align}
R^* = \eta_{\text{th}} \beta^* I^* J^*,
\label{eqn:R}
\end{align}
where $\eta_{\text{th}}\in[0,1]$ describes the fraction of the absorbed optical power that is converted into heat,
$\beta^*$ is the optical attenuation coefficient, 
and $I^*$ is the light intensity.  The absorption of light 
through a material
leads to a decay in the light intensity, which
can be described using the Beer--Lambert law.  
The differential form of the Beer--Lambert law in the current
configuration is given by
\begin{align}
\nx^* I^* \cdot \mathbf{d} = -\beta^* I^*,
\label{eqn:bl_current}
\end{align}
where $\nx^*$ is the gradient expressed in terms of Eulerian
coordinates $x_i$ and $\mathbf{d}$ is the direction
of light propagation.  We assume that the direction of the laser 
is perpendicular to the undeformed upper surface of the beam.
We further assume that the bending of the beam is sufficiently
small so that no refraction occurs.  As a result, we can set
$\mathbf{d} = \mathbf{e}_2$.  
In Lagrangian coordinates, 
the Beer--Lambert law \eqref{eqn:bl_current} is
given by
\begin{align}
\left(\mathbf{F}^*\right)^{-T}\nabla^* I^* \cdot \mathbf{e}_2 = -\beta^* I^*.
\label{eqn:dim_bl}
\end{align}
The prefactor of $(\mathbf{F}^*)^{-T}$ on the left-hand side of
\eqref{eqn:dim_bl} captures the deformation of the path of
light propagation.


\subsection{Boundary and initial conditions}

We consider two sets of thermomechanical boundary conditions.
In the first, the longitudinal ends of the beam are assumed to be free, allowing the beam to freely deform in response to
localised heating (Fig.~\ref{fig:Schematic}~(a)).  At
the ends of the beam, we impose stress-free conditions and
Newton's law of cooling, leading to
\begin{subequations}\label{bc:free}
\begin{align}
    \mathbf{P}^*\mathbf{N}=\mathbf{0}\wat X_1^*=\pm\frac{L^*}{2}, \\
    \mathbf{Q}^*\cdot\mathbf{N}=H^*\LR{\theta^*-\theta_0^*}\mathcal{S}(\mathbf{F}^*) \wat X_1^*=\pm\frac{L^*}2,
\end{align}
\end{subequations}
where $\mathbf{N}$ is the unit normal vector to the boundary, 
$H^*$ is the heat transfer coefficient between the material and the surrounding environment, and
$\mathcal{S} = J^*|\LR{\mathbf{F}^*}^{-T}\mathbf{N}|$ is a dimensionless factor that accounts for the 
difference between area elements in the reference and current
configurations.
We refer to \eqref{bc:free} as free boundary conditions. In the second set of boundary conditions, we assume that the beam is shortened by an amount $\Delta L^*/2$ on each longitudinal end and transverse motion is suppressed.  In addition, the temperature is assumed to be
fixed at the ambient temperature,  leading to
\begin{subequations}\label{bc:clamped}
\begin{align}
    u_1^* = \mp\frac{\Delta L^*}{2}\wat X^*_1=\pm\frac{L^*}2,
    \\
    u_\alpha^* = 0 \wat X^*_1=\pm\frac{L^*}2, 
    \\
    \theta^*=\theta^*_0 \wat X_1^*=\pm\frac{L^*}{2}.
\end{align}
\end{subequations}
We refer to \eqref{bc:clamped} as clamped boundary conditions. Both sets of boundary conditions are combined with
stress-free and Newton-cooling conditions on the transverse
surfaces:
\begin{subequations}
\begin{align}
    \mathbf{P}^*\mathbf{N}=\mathbf{0} \wat X^*_\alpha=\pm\frac{h^*}{2},
    \\
    \mathbf{Q}^*\cdot\mathbf{N}=H^*\LR{\theta^*-\theta_0^*}\mathcal{S}(\mathbf{F}^*) \wat X_\alpha^* =\pm\frac{h^*}2,
\end{align}
\end{subequations}
Mapping the incident light intensity \eqref{eqn:I_0} to 
Lagrangian coordinates leads to the boundary condition
\begin{align}
I^* = \frac{P^*}{\pi (w^*)^2} \exp\left(-\frac{(X_1^* - X_0^* + u_1^*)^2 + (X_3^* + u_3^*)^2}{(w^*)^2}\right)
\wat X_2^* = -\frac{h^*}{2}.
\end{align}
The initial temperature of the beam is assumed to be equal to the
ambient temperature,
\begin{equation}
    \theta^* = \theta_0^* \wat t^*=0.
\end{equation}

\section{Non-dimensionalisation}\label{Scaling}

Lagrangian spatial variables are rescaled as
$X_1^* = L^*X_1$ and $X_\alpha^* =h^*X_\alpha$, 
where $X_1$ and $X_\alpha$ are non-dimensional coordinates.
According to this rescaling, $X_i \in [-1/2, 1/2]$ for $i=1,2,3$. 
The dimensionless gradient operator, $\nabla$, is given by
\begin{equation}
    \nabla = \LR{\delta\pdv{}{X_1},\pdv{}{X_2},\pdv{}{X_3}},
\end{equation}
where $\delta=h^*/L^* \ll 1$ is the small
aspect ratio of the beam. We introduce dimensionless variables
\begin{align}
\begin{split}
t = \frac{t^*}{t^*_{\text{diff}}},
\quad
\mathbf{u} = \frac{\mathbf{u}^*}{h^*},
\quad
\mathbf{x} = \frac{\mathbf{x}^*}{h^*},
\quad
\mathbf{P} = \frac{\mathbf{P}^*}{E^*},
\quad
p = \frac{p^*}{E^*},
\\
I = \frac{\pi (w^*)^2 I^*}{P^*},
\quad
\theta = \frac{\theta^* - \theta^*_0}{\Delta \theta^*},
\quad
\eta = \frac{\theta_0^*\eta^*}{\Delta \theta^* c_\theta^*},
\quad 
\mathbf{K} = \frac{\mathbf{K}^*}{k^*},
\end{split}
\end{align}
where $t^*_\text{diff} = \rho_0^* c_\theta^* (h^*)^2 / k^*$
is the transverse thermal diffusion time
and 
$\Delta \theta^*$ is the characteristic temperature
increase, found by 
balancing the heat supply with the transverse thermal diffusion to give
\begin{equation}\label{Delta theta}
    \Delta\theta^*=
    \frac{\eta_{\text{th}}\beta^*P^*\LR{h^*}^2}{k^*\pi \LR{w^*}^2}.
\end{equation}
Based on experimental data~\cite{Light, wu2025, kim2024}, we expect the relative temperature change to be small, $\varepsilon = \Delta \theta^* / \theta_0^* \ll 1$.

With this scaling, the non-dimensional balance of linear momentum,
balance of angular momentum, incompressibility condition, entropy balance, and Beer--Lambert law are given by
\begin{subequations}
\begin{align}
    \tau^2 \pdv[2]{\mathbf{u}}{t} &= \nabla \cdot\mathbf{P}, 
    \label{lin mom} 
    \\
    \mathbf{P}\mathbf{F}^T&=\mathbf{F}\mathbf{P}^T,
    \label{ang mom}
    \\
    J &= \det \mathbf{F}_\theta,
    \\
    \LR{1 + \varepsilon \theta}\pdv{\eta}{t}&=\nabla\cdot\LR{\mathbf{K}\nabla\theta}+J I ,
    \label{heat eq}
    \\
    \mathbf{F}^{-T} \nabla I \cdot \mathbf{e}_2 &= -\beta I.
    \label{bl}
\end{align}
\end{subequations}
where $\tau = t^*_\text{e} / t^*_\text{diff}$ represents the ratio of the elastic time scale,
$t^*_\text{e} = h^* / \sqrt{E^* / \rho_0^*}$, to the 
time scale of thermal diffusion $t^*_\text{diff}$;
$\mathbf{F}=\mathbf{I}+\nabla\mathbf{u}$ is the deformation
gradient tensor; $J = \det \mathbf{F}$;
$\mathbf{K} = J \mathbf{F}^{-1}\mathbf{F}^{-T}$ is the
dimensionless thermal conductivity tensor;
and $\beta = h^* \beta^*$ is the dimensionless attenuation
coefficient.
The stress tensor is given by
\begin{equation}\label{con eq}
    \mathbf{P}=\frac{2}{3}\left(\mathbf{F}\mathbf{C}_\theta^{-1}-\mathbf{F}^{-T}\right)-p\mathbf{F}^{-T},
\end{equation}
where $\mathbf{C}_\theta = \mathbf{F}_\theta^T \mathbf{F}_\theta$,
\begin{equation}
    \mathbf{F}_\theta=\text{diag}\LR{e^{ \gamma_1\theta},e^{\gamma_2\theta},e^{\gamma_3\theta}},
\end{equation}
and $\gamma_i = \gamma_i^* \Delta \theta^*$ characterise the thermal strains along the principal axes of the beam.  
The non-dimensional entropy can be written as
\begin{align}\label{entropy}
        \eta = \varepsilon^{-1} \ln\left(1 + \varepsilon \theta\right)
        -
        \frac{\mathcal{E}}{3}\pdv{}{\theta}\left[\left(\tr{\mathbf{C}\mathbf{C}_\theta^{-1}}-3
        \right)\right]
        -\mathcal{E}p\mathbf{F}_\theta:\pdv{\mathbf{F}_\theta}{\theta}
\end{align}
where $\mathcal{E} = E^* / (\varepsilon\rho_0^*c_\theta^*\Delta\theta^*)$
is a dimensionless Young's modulus.

\subsection{Non-dimensional boundary and initial conditions}

The non-dimensional free boundary conditions are
\begin{subequations}\label{BCs non dim free}
\begin{align}
    \mathbf{P}\mathbf{N} =\mathbf{0}\wat X_1=\pm\frac{1}{2}, \\
    -\mathbf{K}\nabla \theta \cdot \mathbf{N} = \Bi\, \theta\,\mathcal{S}(\mathbf{F}) \wat X_1=\pm\frac{1}{2},
\end{align}
\end{subequations}
where $\Bi = H^*h^*/k^*$ is the Biot number. 
The non-dimensional clamped boundary conditions are
\begin{subequations}\label{BCs non dim clamped}
\begin{align}
    u_1 = \mp\frac{\Delta L}{2}\wat X_1=\pm\frac{1}2,
    \\
    u_\alpha = 0 \wat X_1=\pm\frac{1}2, 
    \\
    \theta=0 \wat X_1=\pm\frac{1}{2},
\end{align}
\end{subequations}
where $\Delta L = \Delta L^* / h^*$.
At the transverse faces of the beam, the boundary conditions are
\begin{subequations}
\begin{align}
    \mathbf{P}\mathbf{N} = \mathbf{0}\wat X_\alpha=\pm\frac12, \\
    -\mathbf{K}\nabla\theta\cdot\mathbf{N} = \Bi\,\theta\, \mathcal{S}(\mathbf{F})\wat X_\alpha=\pm\frac12.
    \label{bc:newton}
\end{align}
\end{subequations}
The boundary condition for the dimensionless laser intensity is
\begin{align}
I = \exp\left(-\frac{(X_1 - X_0 + \delta u_1)^2}{\delta^2 w^2} -\frac{(X_3 + u_3)^2}{w^2}\right)
\wat X_2 = -\frac{1}{2},
\end{align}
where $w = w^* / h^*$ and $X_0 = X_0^* / L^*$.
Finally, the initial condition on the temperature is 
\begin{equation}
    \theta=0 \wat t=0.
\end{equation}

\section{Asymptotic reduction}\label{Asymptotics}

The dimensionless governing equations are reduced by
leveraging the slenderness of the beam and taking the asymptotic
limit $\delta \to 0$.  
As thermal diffusion is much slower than elastic relaxation, $\tau\ll1$.  We thus consider the distinguished limit in which $\tau = \delta^2\epsilon$, which ensures that inertia is captured in the reduced model.  
We work in the limit of small thermal strains and small
end shortening (when applied) so that
$\gamma_i = \delta \Gamma_i$ and $\Delta L = \delta \lambda$.
We treat $\epsilon$, $\Gamma_i$, $\lambda$, and all
other non-dimensional parameters as being $O(1)$ when
$\delta \to 0$.



To simplify the mechanics, the analysis is restricted to
cases when the cross-section of the beam, to leading order,
does not 
rotate, corresponding to twisting, is not sheared, and is
not stretched.  Consequently, the 
transverse displacements will have the asymptotic form
$u_\alpha(\mathbf{X}, t) \sim u_\alpha^{(0)}(X_1, t)$.
We follow classical beam and rod theories \cite{howell2008} and 
assume that longitudinal displacements are small,
$u_1 = O(\delta)$.


The form of the boundary condition \eqref{eqn:I_0}
shows that the incident light intensity will be exponentially
small except in a localised region near the irradiation point where
$|X_1 - X_0| = O(\delta)$.  Thus, the domain is decomposed into an outer
region where $|X_1 - X_0| = O(1)$ and an inner region where
$|X_1 - X_0| = O(\delta)$.  Matched asymptotic expansions are used to 
reduced the governing equations in the outer regions and derive
asymptotic consistent jump conditions by resolving the solution
in the inner region.

\subsection{Outer problems}\label{Outer problem}

In the outer region, where $|X_1-X_0|=O(1)$, 
the displacements are expanded as
$u_1(\mathbf{X},t) = \delta u^{(0)}_{1,\pm}(\mathbf{X},t)+O\LR{\delta^2}$
and $u_\alpha(\mathbf{X},t) =u_{\alpha,\pm}^{(0)}(X_1,t)+O(\delta)$,
where the $\pm$ subscript refers to the left and right outer regions, defined by $X_1 < X_0$ and $X_1 > X_0$, respectively.  The deformation gradient tensor has the
asymptotic form
$\mathbf{F}=\mathbf{I}+\delta\mathbf{F}_\pm^{(1)}+O\LR{\delta^2}$, where
\begin{equation}
    \mathbf{F}^{(1)}_\pm = \pdv{u_{1,\pm}^{(0)}}{X_\alpha}\eoe1\alpha+\pdv{u_{\alpha,\pm}^{(0)}}{X_1}\eoe\alpha1+\pdv{u_{\alpha,\pm}^{(1)}}{X_\beta}\eoe\alpha\beta,
    \label{outer:F1}
\end{equation}
and summation over repeated Greek indices is implied.
From the non-dimensional Beer--Lambert law \eqref{bl}, we then deduce that
the light intensity is exponentially small, so there is no heat supply in the
outer regions.  Moreover, the dimensionless entropy \eqref{entropy} can be
expanded as $\eta = \theta_{\pm}^{(0)} + O(\delta)$, from which we deduce that
$\theta_{\pm}^{(0)} \equiv 0$ in the outer regions.  Thus, the beam is isothermal
and behaves as an elastic material to leading order.

The elastic problem for the beam is reduced by first
noticing that $P_{11}=O\LR{\delta^2}$ from the
constitutive equation \eqref{con eq} and the form of the deformation gradient tensor \eqref{outer:F1}.  As a result, 
$p = O(\delta^2)$ as well.  
Moreover, by balancing terms in \eqref{lin mom}, we
deduce that $P_{1\alpha},\,P_{\alpha 1} = O(\delta^3)$
and $P_{\alpha\beta} = O(\delta^4)$.  
Thus, the stresses and Lagrange muliplier are expanded as
\begin{subequations}
\begin{align}
    (P_{11},p) &= \delta^2\LR{P_{11,\pm}^{(0)},p_{\pm}^{(0)}}+O\LR{\delta^3},\\
    (P_{1\alpha},P_{\alpha1,\pm})&=\delta^3\LR{P^{(0)}_{1\alpha,\pm},P^{(0)}_{\alpha1,\pm}}+O\LR{\delta^4},\\
    P_{\alpha\beta} &=\delta^4P_{\alpha\beta,\pm}^{(0)}+O\LR{\delta^5}.
\end{align}
\end{subequations}
In the outer region, the leading-order balance of linear momentum in the longitudinal and transverse directions are, respectively,
\begin{subequations}
\begin{align}\label{longitudinal stress balance}
    0 &= \pdv{P_{11,\pm}^{(0)}}{X_1}+\pdv{P_{1\beta,\pm}^{(0)}}{X_\beta}, \\
    \label{transverse stress balance}
    \epsilon^2\pdv[2]{u_{\alpha,\pm}^{(0)}}{t} &= \pdv{P_{\alpha1,\pm}^{(0)}}{X_1}+\pdv{P_{\alpha\beta,\pm}^{(0)}}{X_\beta}.
\end{align}
\end{subequations}
Integrating these equations over the beam cross section, and using the traction-free boundary conditions on the transverse surfaces, leads to
\begin{subequations}
\begin{align}\label{P11 eq}
    \pdv{\B{P_{11,\pm}^{(0)}}}{X_1}&=0, \\
    \epsilon^2\pdv[2]{u_{\alpha,\pm}^{(0)}}{t} &= \pdv{\B{P_{\alpha1,\pm}^{(0)}}}{X_1}, \label{transverse momentum balance}
\end{align}
\end{subequations}
where the cross-sectional average of a quantity $f$ is defined as
\begin{align}
\bar{f} = \int_{-1/2}^{1/2}\int_{-1/2}^{1/2} f\,\mathrm{d} X_2 \mathrm{d} X_3.
\label{eqn:average}
\end{align}
In \eqref{transverse momentum balance}, we omit the bar over $u_{\alpha,\pm}^{(0)}$ as it does not depend on the transverse coordinates. 
Thus, the averaged longitudinal stress $\B{P_{11,\pm}^{(0)}}$ is independent of $X_1$ in both outer regions.  
The constitutive equation \eqref{con eq} gives at $O(\delta)$
\begin{equation}
    \pdv{u_{1,\pm}^{(0)}}{X_\alpha}+\pdv{u_{\alpha,\pm}^{(0)}}{X_1} = 0.
\end{equation}
Since $u_{\alpha,\pm}^{(0)}$ is independent of $X_\alpha$, these can be integrated to determine the
longitudinal displacement in terms of the transverse displacements,
\begin{equation}\label{u 1 identity}
    u_{1,\pm}^{(0)}(X_1,X_\alpha,t) = U_\pm(X_1,t)-X_\alpha\pdv{u_{\alpha,\pm}^{(0)}}{X_1},
\end{equation}
where $U_\pm$ is the longitudinal displacement of the centreline of the beam. 

To close the problem for the transverse displacements,
expressions for $\B{P_{\alpha1,\pm}^{(0)}}$ are required.
The $\eoe1\alpha$ term in the $O(\delta)$ contribution to the balance of angular momentum \eqref{TE:ang mom dim} is given by
\begin{equation}
    P_{1\alpha,\pm}^{(0)}-P_{\alpha1,\pm}^{(0)} = P_{11,\pm}^{(0)}\pdv{u_{1,\pm}^{(0)}}{X_\alpha}.
\end{equation}
Taking the average of this and making use of the expression for $u_{1,\pm}^{(0)}$ in \eqref{u 1 identity} gives
\begin{equation}\label{angular momentum relation}
    \B{P_{\alpha1,\pm}^{(0)}} = \B{P_{1\alpha,\pm}^{(0)}}+\B{P_{11,\pm}^{(0)}}\pdv{u_{\alpha,\pm}^{(0)}}{X_1}.
\end{equation}
Multiplying the longitudinal stress balance \eqref{longitudinal stress balance} by $X_\alpha$, averaging, and using the traction-free boundary conditions gives the moment-shear relation
\begin{equation}\label{moment relation}
    \B{P_{1\alpha,\pm}^{(0)}} = \pdv{\B{X_\alpha P_{11,\pm}^{(0)}}}{X_1}.
\end{equation}
Thus, all that remains is to calculate $P_{11,\pm}^{(0)}$.
    %

Expanding the incompressibility condition \eqref{incompressibility condition} gives
\begin{equation}
    \tr{\mathbf{F}_\pm^{(1)}}=0\wand \tr{\mathbf{F}_\pm^{(2)}}=\frac12\tr{\LR{\mathbf{F}_\pm^{(1)}}^2}.
\end{equation}
Thus, taking the trace of the $O(\delta^2)$ contribution to the constitutive relation 
\eqref{con eq} yields
\begin{equation}
    p_\pm^{(0)} = -\frac13P_{11,\pm}^{(0)}.
\end{equation}
Eliminating the Lagrange multiplier from the $\eoe11$ element of the constitutive equation gives, at $O(\delta^2)$,
\begin{equation}
    P_{11,\pm}^{(0)} = \pdv{U_\pm}{X_1}-X_\beta\pdv[2]{u_{\beta,\pm}^{(0)}}{X_1}+\frac12\pdv{u_{\beta,\pm}^{(0)}}{X_1}\pdv{u_{\beta,\pm}^{(0)}}{X_1},
    \label{outer:P11}
\end{equation}
The three terms on the right-hand side correspond to centreline stretching, bending, and a nonlinearvon K\'arm\'an strain 
capturing geometric nonlinearities.
Averaging \eqref{outer:P11} over the cross-section gives
\begin{equation}\label{P11 outer expression}
    \B{P_{11,\pm}^{(0)}} = \pdv{U_\pm}{X_1}+\frac12\pdv{u_{\beta,\pm}^{(0)}}{X_1}  \pdv{u_{\beta,\pm}^{(0)}}{X_1},
\end{equation}
as $\B{X_\beta}=0$. Multiplying \eqref{outer:P11} by $X_\alpha$ and averaging gives
\begin{equation}\label{moment expression}
    \B{X_\alpha P_{11,\pm}^{(0)}} = -\frac1{12}\pdv[2]{u_{\alpha,\pm}^{(0)}}{X_1}.
\end{equation}
Substituting \eqref{moment expression} into \eqref{moment relation} gives
\begin{equation}
    \B{P_{1\alpha,\pm}^{(0)}} = -\frac1{12}\pdv[3]{u_{\alpha,\pm}^{(0)}}{X_1}.
\end{equation}
The angular-momentum relation \eqref{angular momentum relation} then gives
\begin{equation}\label{Shear stress outer}
    \B{P_{\alpha1,\pm}^{(0)}} = -\frac1{12}\pdv[3]{u_{\alpha,\pm}^{(0)}}{X_1}+\B{P_{11,\pm}^{(0)}}\pdv{u_{\alpha,\pm}^{(0)}}{X_1}.
\end{equation}
Substituting this into the transverse momentum balance \eqref{transverse momentum balance} yields
\begin{equation}
    \epsilon^2\pdv[2]{u_{\alpha,\pm}^{(0)}}{t}=-\frac1{12}\pdv[4]{u_{\alpha,\pm}^{(0)}}{X_1}+\B{P_{11,\pm}^{(0)}}\pdv[2]{u_{\alpha,\pm}^{(0)}}{X_1}.
    \label{b1}
\end{equation}
This equation holds in both outer regions.  Equation \eqref{b1}
is a standard form of beam equation~\cite{audoly2010,Antman2005}; however, geometric nonlinearities are accounted for through the longitudinal stress resultant $\B{P_{11,\pm}^{(0)}}$ given by \eqref{P11 outer expression}.

\subsection{Inner problems}\label{Inner problem}

In the inner region, we define the inner variable as $\XM=\delta^{-1}(X_1-X_0)$. The displacements are scaled the same as in the outer regions,
$u_1 = \delta u_{1,\i}^{(0)}(\XM, X_2, X_3,t) + O(\delta^2)$ and
$u_{\alpha} = u_{\alpha, \i}^{(0)}(\XM, t) + O(\delta)$,
but the deformation gradient tensor now has the asymptotic form $\mathbf{F}=\mathbf{F}_{\i}^{(0)}+\delta\mathbf{F}_{\i}^{(1)}+O(\delta^2)$, where
\begin{equation}\label{F_i 0}
    \mathbf{F}_{\i}^{(0)} = \mathbf{I}+\pdv{u_{\alpha,\i}^{(0)}}{\XM}\eoe\alpha1.
\end{equation}
From the balance of linear and angular momentum, we deduce that all components of the stress tensor must now be the same order 
of magnitude. 
Therefore, the stress tensor is expanded as
$\mathbf{P} = \delta^2 \mathbf{P}_{\i}^{(0)} + \delta^3 \mathbf{P}_{\i}^{(1)} + O(\delta^4)$
and Lagrange multiplier is written as $p = \delta^2 p_{\i}^{(0)} + O(\delta^3)$. The laser intensity and temperature are expanded  as $I_{\i}=I_{\i}^{(0)}+O(\delta)$ and $\theta = \theta_{\i}^{(0)} + O(\delta)$. The thermal distortion is expanded as
$\mathbf{F}_\theta = \mathbf{I}+\delta\mathbf{F}_\theta^{(1)}\theta_{\i}^{(0)}+O\LR{\delta^2}$ where $\mathbf{F}_\theta^{(1)} = \text{diag}(\Gamma_1,\Gamma_2,\Gamma_3)$.
Material incompressibility implies that 
$J = \det \mathbf{F}_{\theta} = 1 + \delta \theta_{\i}^{(0)} \tr{\mathbf{F}_\theta^{(1)}} + O(\delta^2)$.  From the
$O(1)$ contributions to the constitutive relation
\eqref{con eq}, we find that $\mathbf{F}_{\i}^{(0)} - [\mathbf{F}_{\i}^{(0)}]^{T} = \mathbf{0}$, which implies
that
\begin{align}
\pdv{u_{\alpha,\i}^{(0)}}{\XM} = 0.
\label{u_alpha inner}
\end{align}
Thus, the transverse displacements, to leading order, are
spatially uniform across the inner region.  We therefore let
$u_{\alpha, \i}^{(0)} = u_{\alpha, \i}^{(0)}(t)$. Moreover,
the leading-order contribution to the deformation gradient tensor is equal to the identity tensor, 
$\mathbf{F} = \mathbf{I} + O(\delta)$.  The balance of linear
momentum then implies that the leading-order contribution to the stress tensor is symmetric, $\mathbf{P}_{\i}^{(0)} = [\mathbf{P}_{\i}^{(0)}]^T$.

\subsubsection{Inner thermal problem}\label{Inner Thermal problem}

The inner thermal problem is formulated by first calculating
the light intensity.  Using the fact that $\mathbf{F} \sim \mathbf{I}$ in the Beer--Lambert law \eqref{bl} and integrating,
we obtain
\begin{equation}\label{I_i}
    I_{\i}^{(0)} = \exp\left(-\frac{\XM^2+\LR{X_3 + u_{3, \i}^{(0)}(t)}^2}{w^2} \right)\exp\LR{-\beta\LR{X_2+\frac12}}.
\end{equation}
The expression for the entropy \eqref{entropy} implies that $\eta_{\i}=\theta_{\i}^{(0)}+O(\delta)$ in the inner region.
Moreover, $J \sim 1$ and $\mathbf{K} \sim \mathbf{I}$.
Thus, the leading-order thermal problem found from
\eqref{heat eq} is 
\begin{equation}\label{inner thermal problem eq}
    \pdv{\theta_{\i}^{(0)}}{t} = \pdv[2]{\theta^{(0)}_{\i}}{\XM}+\frac{\p^2 \theta_{\i}^{(0)}}{\p X_\alpha \p X_\alpha}+I_{\i}^{(0)},
\end{equation}
where $I_{\i}^{(0)}$ is given in \eqref{I_i}.  
The transverse boundary conditions for \eqref{inner thermal problem eq} are obtained from the $O(1)$ contributions to
the Newton cooling conditions \eqref{bc:newton},
\begin{equation}
    \pdv{\theta_{\i}^{(0)}}{X_\alpha}\pm\Bi\,\theta_{\i}^{(0)} = 0\wfor X_\alpha =\pm\frac12.
\end{equation}
The longitudinal boundary conditions are obtained by matching
to the outer solution to find
\begin{equation}
    \theta_{\i}^{(0)}\to 0 \was \XM\to\pm\infty.
\end{equation}
Finally, the initial condition for \eqref{inner thermal problem eq} is
\begin{equation}
    \theta_{\i}^{(0)}=0 \wat t=0.
    \label{eqn:inner_ic}
\end{equation}
By treating $u_{3, \i}^{(0)}(t)$ as a known, time-dependent quantity,
the inner thermal problem \eqref{inner thermal problem eq}--\eqref{eqn:inner_ic} becomes linear and can be solved using
standard methods.

Once the inner thermal problem is solved, three quantities
can be computed that are required for the inner mechanical
problem.  These are the mean inner temperature,
\begin{equation}\label{Theta 0 def}
   \Theta = \int_{-\infty}^\infty\int_{-1/2}^{1/2}\int_{-1/2}^{1/2} \theta_{\i}^{(0)}\text{d}X_2\text{d}X_3\text{d}\XM,
\end{equation}
and the first transverse thermal moments
\begin{equation}\label{Theta a def}
    M_{\alpha} = \int_{-\infty}^\infty\int_{-1/2}^{1/2}\int_{-1/2}^{1/2} X_\alpha\theta_{\i}^{(0)}\text{d}X_2\text{d}X_3\text{d}\XM.
\end{equation}
Due to the symmetry of the problem about $X_3$, we find 
that $M_3 \equiv 0$.

\subsubsection{Inner mechanical problem}


In the inner region, the leading-order balance of linear momentum in the longitudinal and transverse directions is, respectively,
\begin{align}
    \pdv{P_{11,\i}^{(0)}}{\XM}+\pdv{P_{1\beta,\i}^{(0)}}{X_\beta}&=0,\label{longitudinal stress balance inner}\\
    \pdv{P_{\alpha1,\i}^{(0)}}{\XM}+\pdv{P_{\alpha\beta,\i}^{(0)}}{X_\beta}&=0.
\end{align}
Taking the cross-sectional average of these and applying the traction-free boundary conditions on the transverse surfaces
leads to
\begin{align}
    \pdv{\B{P_{11,\i}^{(0)}}}{\XM}&=0, \label{inner:P11} \\
    \pdv{\B{P_{\alpha 1,\i}^{(0)}}}{\XM}&=0.\label{Pa1 0 inner}
\end{align}
In the outer region, $P_{\alpha 1} = O(\delta^3)$.  
Therefore, by integrating \eqref{Pa1 0 inner} and matching,
we obtain $P_{\alpha 1, \i}^{(0)} \equiv 0$.  
Multiplying \eqref{longitudinal stress balance inner} by $X_\alpha$ and integrating gives
\begin{equation}\label{Inner moment}
    \pdv{\B{X_\alpha P_{11,\i}^{(0)}}}{\XM}=\B{P_{1\alpha,\i}^{(0)}} = \B{P_{\alpha 1,\i}^{(0)}} = 0.
\end{equation}
where the second equality arises from using  the symmetry of the leading-order stress tensor. 
At $O(\delta^3)$ in the inner region, the balance of linear momentum in the transverse direction is 
\begin{equation}
    \pdv{P_{\alpha1,\i}^{(1)}}{\XM}+\pdv{P_{\alpha\beta,\i}^{(1)}}{X_\beta}=0,
\end{equation}
taking the average of this over the cross-section and making use of the traction-free boundary conditions yields
\begin{equation}\label{inner shear balance}
    \pdv{\B{P_{\alpha1,\i}^{(1)}}}{\XM}=0,
\end{equation}
which implies that the leading-order shear stresses in the outer region are continuous over the inner region.
The $O(\delta)$ contributions to the constitutive equation in the inner region lead to
\begin{equation}
    0=\frac23\LRB{\mathbf{F}_{\i}^{(1)}+\LR{\mathbf{F}_{\i}^{(1)}}^T-\LR{\mathbf{F}_\theta^{(1)}+\LR{\mathbf{F}_\theta^{(1)}}^T}\theta_{\i}^{(0)}}.
\end{equation}
The $\eoe11$ component of the above gives
\begin{equation}\label{P11 inner identity}
    \pdv{u_{1,\i}^{(0)}}{\XM}-\Gamma_1\theta_{\i}^{(0)}=0,
\end{equation}
This relationship is one of the key mechanisms through which the inner temperature field enters the elastic problem.



\subsection{Jump conditions across the inner region}

Asymptotically consistent jump conditions
for the outer problem are derived by asymptotic matching of
the inner and outer solutions.
For a quantity $f$ in the outer region, 
its jump across the inner region is defined as
\begin{equation}
    \LRB{f_\pm(X_1)}_-^+=\lim_{X_1\to X_0^+}f_+(X_1)-\lim_{X_1\to X_0^-}f_-(X_1).
\end{equation}
In our case, asymptotic matching requires
\begin{align}
\lim_{\XM \to \pm \infty} f^{(0)}_{\i} = \lim_{X_1 \to X_{0}^{\pm}} f^{(0)}.
\end{align}
%
%
%
Integrating \eqref{u_alpha inner} across the inner region
and matching leads to 
\begin{align}
\LRB{u_{\alpha,\pm}^{(0)}}_-^+=0,
\end{align}
implying the (outer) transverse displacements are continuous.
The quantity $u_{3, \i}^{(0)}(t)$ appearing in \eqref{I_i} is simply
obtained by evaluating the transverse displacement at
$X_0$ to give 
\begin{align}
u_{3, \i}^{(0)}(t) = u_{3, +}^{(0)}(X_0, t) = u_{3, -}^{(0)}(X_0, t).
\label{U3}
\end{align}
By taking the cross-sectional average of \eqref{P11 inner identity}, 
integrating across the inner region, and matching to
the average of \eqref{u 1 identity} leads to
\begin{equation}\label{U jump}
    \LRB{U_\pm}_-^+ = \Gamma_1\Theta,
\end{equation}
where we have made use of \eqref{Theta 0 def}.  Thus,
there is a jump in the longitudinal displacement due to
localised thermal expansion.  


A jump condition on the gradient of the transverse displacements
can be obtained by multiplying \eqref{P11 inner identity} by
$X_\alpha$, taking the cross-sectional average, integrating
over the inner region, and matching using \eqref{u 1 identity}
to obtain
\begin{equation}\label{slope jump}
    \LRB{\pdv{u_{\alpha,\pm}^{(0)}}{X_1}}_-^+ = -12\Gamma_1 M_\alpha.
\end{equation}
Therefore, the transverse thermal moments $M_\alpha$ act 
as local hinges that create kinks in the beam.



Integrating the mean stress balance 
\eqref{inner:P11} across the inner region and matching shows
that the mean longitudinal stresses $\B{P_{11,\pm}^{(0)}}$ are continuous across the inner region.  Moreover, since
$\B{P_{11, \pm}}$ are also spatially uniform in the outer
regions, it follows that the mean longitudinal stress is
uniform across the entire beam,
\begin{align}
\B{P_{11,+}^{(0)}} = \B{P_{11,-}^{(0)}} \equiv \B{P_{11}^{(0)}}.
\label{jump:P11}
\end{align}
Similarly, integrating the moment balance \eqref{Inner moment} 
across the inner region shows that the moments are continuous;
matching to \eqref{moment expression} then produces
a jump condition on the second derivative of the transverse
displacement,
\begin{equation}
    \LRB{\pdv[2]{u_{\alpha,\pm}^{(0)}}{X_1}}_-^+ =0.
\end{equation}
Finally, integrating \eqref{inner shear balance} across the
inner region shows that the mean shear stresses
are continuous across the inner region.  Matching to the mean
shear stress in the outer regions \eqref{Shear stress outer}
gives
\begin{equation}
    \LRB{-\frac1{12}\pdv[3]{u_{\alpha,\pm}^{(0)}}{X_1}+\B{P_{11,\pm}^{(0)}}\pdv{u_{\alpha,\pm}^{(0)}}{X_1}}_-^+=0.
\end{equation}
By rearranging and making use of \eqref{slope jump} and 
\eqref{jump:P11}, we obtain
\begin{equation}
    \LRB{\pdv[3]{u_{\alpha,\pm}^{(0)}}{X_1}}_-^+ = -144\Gamma_1\B{P_{11}^{(0)}}M_\alpha.
\end{equation}

\subsection{Boundary conditions for the outer mechanical problem}

To close the outer problem, mechanical boundary conditions
at the longitudinal ends of the beam ($X_1 = \pm 1/2)$
are required.  Given that the thermoelastic model reduces to
an elastic model in the outer region, we simply state the
well-established clamped and free boundary conditions here. However, 
in general, boundary conditions for beam models should be derived using
matched asymptotic expansions to resolve any Saint-Venant boundary  
layers~\cite{howell2008, hennessy2024}.


The boundary conditions in the case of clamped ends are given by
\begin{subequations}\label{outer:bc_clamped}
\begin{align}
     u_{\alpha,\pm}^{(0)}=0\wat X_1=\pm\frac12, \\
     \pdv{u_{\alpha,\pm}^{(0)}}{X_1} = 0\wat X_1 = \pm\frac{1}{2}, \\
     U_\pm=\mp\frac\lambda2\wat X_1=\pm\frac12. \label{outer:U_bc}
\end{align}
\end{subequations}
By using \eqref{outer:U_bc}, it possible to determine the
mean longitudinal stress through a longitudinal force balance
(see Appendix~\ref{app:axial}) to find
\begin{align}
    \B{P_{11}^{(0)}} = -\lambda-\Gamma_1\Theta+\frac12
    \int_{-1/2}^{X_0}\pdv{u_{\beta,-}^{(0)}}{X_1}\pdv{u_{\beta,-}^{(0)}}{X_1}\,\mathrm{d}X_1
    +\frac{1}{2}\int_{X_0}^{1/2}\pdv{u_{\beta,+}^{(0)}}{X_1}\pdv{u_{\beta,+}^{(0)}}{X_1}\,\mathrm{d}X_1.
    \label{outer:P11_soln}
\end{align}
When solving the outer problem with clamped boundary conditions,
\eqref{outer:P11_soln} is used in place of \eqref{outer:U_bc}.

The boundary conditions in the case of free ends are given by
\begin{align}
\pdv[2]{u_{\alpha,\pm}^{(0)}}{X_1}=\pdv[3]{u_{\alpha,\pm}^{(0)}}{X_1}=0\wat X_1=\pm\frac12. \label{outer:bc_free}
\end{align}
Moreover, in this case, the mean longitudinal stress is zero throughout the beam,
\begin{align}
\B{P_{11}^{(0)}} \equiv 0.
\end{align}


\section{Further reductions of the model}\label{Reduced model}

Due to the symmetry of the problem about $X_3 = 0$, the conclusion that $M_3 = 0$, and the smallness of the thermal
strains, we find that $u_3^{(0)} \equiv 0$.  Therefore, 
to leading order, the beam
undergoes plane strain in the $X_1$-$X_2$ plane.  
We now 
collect the results of the previous sections into a final reduced model for the 
actuated thermoelastic beam.
To write the model in a clearer form, we introduce the rescaled quantities
\begin{align}
\begin{split}
\W{t} = \frac{t}{\sqrt{12}},
\quad
\W{v}_{\pm} = \sqrt{12}u_{2,\pm}^{(0)},
\quad
\W{\mu}^2=-12\B{P_{11}^{(0)}},
\quad
\W{\lambda}=12\lambda,
\quad 
\W{\Theta}=12\Theta,
\quad 
\W{M} = 12\sqrt{12}M_2.
\end{split}
\end{align}
After these rescalings, the governing equation for the
transverse displacement is
\begin{equation}\label{eq: beam model}
    \epsilon^2\pdv[2]{\W{v}_{\pm}}{\W{t}}+\pdv[4]{\W{v}_{\pm}}{X_1}+\W{\mu}^2\pdv[2]{\W{v}_{\pm}}{X_1}=0.
\end{equation}
The four jump conditions at $X_1 = X_0$ are given by
\begin{subequations}\label{outer_bc_clamped}
\begin{align}
\LRB{\W{v}_{\pm}}_-^+ &= 0,
\\
\LRB{\pdv{\W{v}_{\pm}}{X_1}}_-^+&=-\Gamma_1\W{M}, \label{red:bc_first}
\\
\LRB{\pdv[2]{\W{v}_{\pm}}{X_1}}_-^+&=0,
\\
\LRB{\pdv[3]{\W{v}_{\pm}}{X_1}}_-^+&=\Gamma_1\W{\mu}^2\W{M},
\end{align}
\end{subequations}
where $\W{\Theta}$ and $\W{M}$ are defined in terms of the inner
temperature field $\theta_{\i}^{(0)}$ as
\begin{align}
    \W\Theta &= 12\int_{-\infty}^\infty\int_{-1/2}^{1/2}\int_{-1/2}^{1/2}\theta_{\i}^{(0)}\,\text{d}X_2\text{X}_3\text{d}\XM, \\
    \W{M} &= 12\sqrt{12}\int_{-\infty}^\infty\int_{-1/2}^{1/2}\int_{-1/2}^{1/2} X_2\theta_{\i}^{(0)}\,\text{d}X_2\text{d}X_3\text{d}\XM.
\end{align}
Since $u_{3, \i}^{(0)}\equiv 0$, the transverse displacement no longer
enters expression for the light intensity \eqref{I_i}.  Thus,
the inner thermal problem decouples from the mechanical problem
and can be solved independently.  A series solution for the inner temperature field
is provided in Appendix~\ref{app:thermal}.

In the case of clamped ends, the boundary conditions are given by
\begin{subequations}
\begin{equation}
    \W{v}_{\pm}=\pdv{\W{v}_{\pm}}{X_1}=0\wat X_1=\pm\frac12,
    \label{reduced:bc_clamped}
\end{equation}
and the longitudinal force balance is imposed,
\begin{equation}\label{compcond}
    \W{\mu}^2 = \W{\lambda}+\Gamma_1\W{\Theta}-\frac12
    \int_{-1/2}^{X_0}\left(\pdv{\W{v}_{-}^{(0)}}{X_1}\right)^2\,\mathrm{d}X_1
    -\frac{1}{2}\int_{X_0}^{1/2}\left(\pdv{\W{v}_{+}^{(0)}}{X_1}\right)^2\,\mathrm{d}X_1.
\end{equation}
\end{subequations}
In the case of free ends, the boundary conditions are
\begin{equation}
    \pdv[2]{\W{v}_{\pm}}{X_1}=\pdv[3]{\W{v}_{\pm}}{X_1}=0\wat X_1=\pm\frac12,
    \label{reduced:bc_free}
\end{equation}
and the longitudinal force is zero, $\W{\mu} \equiv 0$. 


For convenience, we also provide the dimensional form of the reduced model.
The equation of motion for the transverse displacement 
$v^* = u_2^*$ is given by
\begin{equation}
    \rho_0^*A^*\pdv[2]{v^*_{\pm}}{t^*}+E^*I_\alpha^*\pdv[4]{v_{\pm}^*}{X_1^*}+N^*\pdv[2]{v_{\pm}^*}{X^*_1}=0,
\end{equation}
where $A^* = (h^*)^2$ is the area of the cross section, $I_\alpha^*=(h^*)^4/12$ is the second area moment for the cross section, and $N^*$ is the longitudinal stress resultant, which
is uniform throughout the beam but can depend on time.
The jump conditions at $X_1^* = X_0^*$ are given by
\begin{subequations}
\begin{align}
    \LRB{v^*_{\pm}}_-^+ &= 0, 
    \\
    \LRB{\pdv{v_{\pm}^*}{X_1^*}}_-^+ &= \gamma_1^*M^*(t^*),
    \\
    \LRB{\pdv[2]{v_{\pm}^*}{X^*_1}}_-^+&=0,
    \\
    \LRB{\pdv[3]{v_{\pm}^*}{X_1^*}}_-^+ &= \frac{\gamma_1^*N^*(t^*)M^*(t^*)}{E^*I_\alpha^*}.
\end{align}
\end{subequations}
The dimensional thermal forcing terms are 
\begin{align}
    \Theta^*(t^*)&=\frac1{A^*}\int_{-\infty}^\infty\int_{A^*}\LR{\theta_{\i}^*-\theta_0^*}\text{d}A^*\text{d}\XM^*,
    \\
    M^*(t^*)&=\frac1{I_\alpha^*}\int_{-\infty}^\infty\int_{A^*} X_{2}^*\LR{\theta_{\i}^*-\theta_0^*}\text{d}A^*\text{d}\XM^*,
\end{align}
where $A^*$ is the cross-sectional area and
$\theta^*_{\i}$ is the temperature distribution in the inner region.

In the case of clamped boundary conditions, the longitudinal stress resultant is given 
by
\begin{equation}
    N^* = \frac{E^*A^*}{L^*}\LRB{\Delta L^*+\Gamma_1^*\Theta^*-\frac12\int_{-L^*/2}^{X_0^*}\left(\pdv{v_{-}^*}{X_1^*}\right)^2\,\text{d}X_1^* - 
    \frac12\int_{X_0^*}^{L^*/2}\left(\pdv{v_{+}^*}{X_1^*}\right)^2\,\text{d}X_1^*}.
\end{equation}
In the case of free boundary conditions, $N^* \equiv 0$.  The dimensional boundary
conditions for the transverse displacements are 
analogous to those in \eqref{reduced:bc_clamped} and \eqref{reduced:bc_free}.

\section{Static analysis}\label{Static analysis}

We now make use of the dimensionless reduced model presented in Sec.~\ref{Reduced model},
with tildes dropped for notational convenience, to study the equilibrium response of a beam under localised heating.  
%
%
%
We focus on two physical situations involving thermoresponsive hydrogel beams. The first examines how localised heating leads to a $V$-shaped deformation in a free gel (Fig.~\ref{fig:Schematic}~(a)). The second examines how localised heating can induce snap-through in 
pre-buckled hydrogel beams that are clamped at the ends (Fig.~\ref{fig:Schematic}~(b)).  The parameters are based on 
the light-actuated PNIPAM gels created by Dai et al.~\cite{Light}, which
have negative coefficients of thermal expansion.  Details of the
parameter values can be found in Appendix~\ref{sec:Parameter values}.

\subsection{$V$-shape deformation of free hydrogel beams}

\begin{figure}
    \centering
    \includegraphics[width=0.95\linewidth]{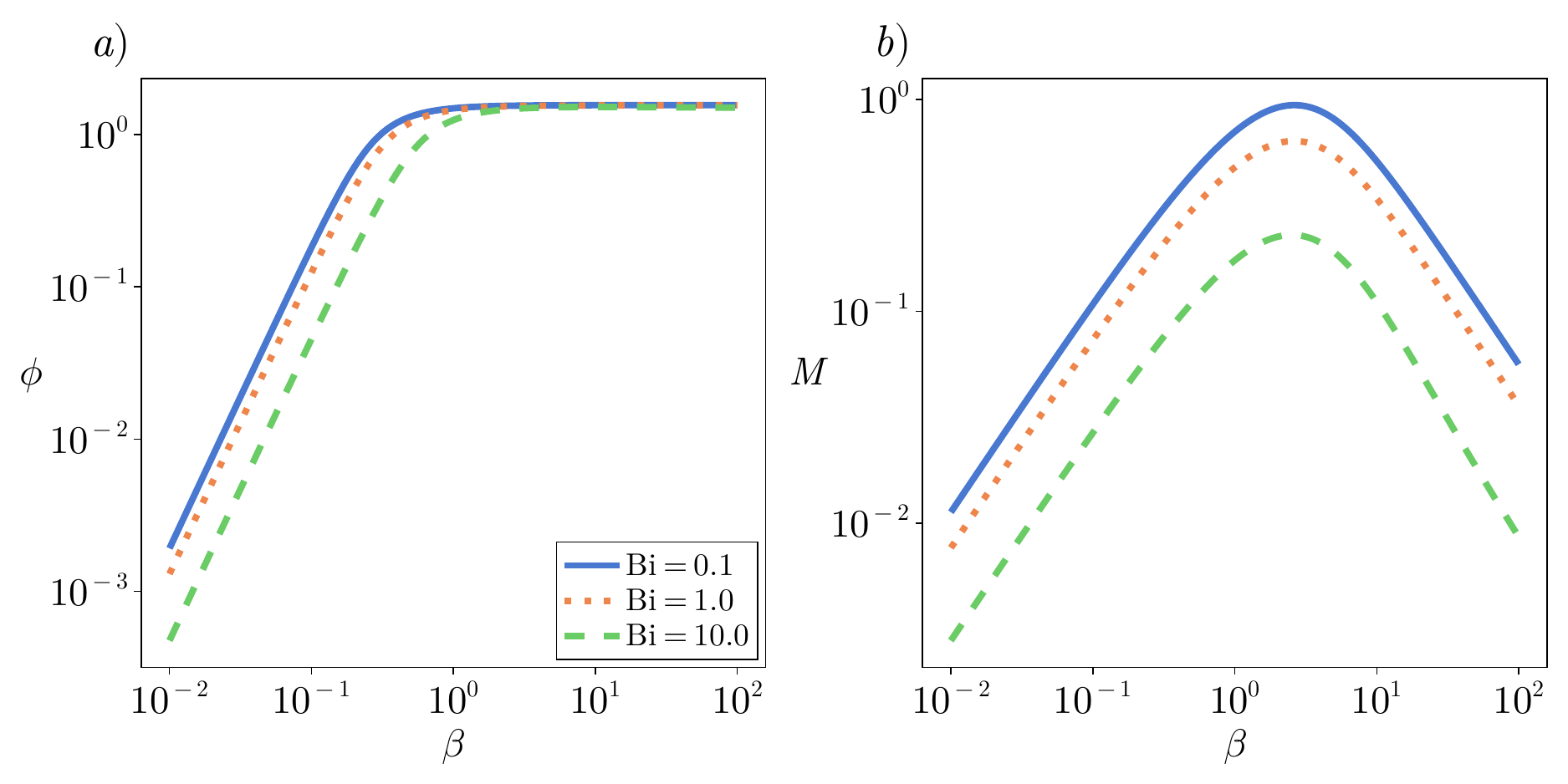}
    \caption{(a) The fold angle $\phi$ of V-shaped beams as a function of the dimensionless optical attenuation coefficient $\beta = \beta^* h^*$ at different values of the Biot number $\Bi = H^* h^* / k^*$.  (b) The dimensionless thermal moment as a function of the attenuation coefficient and Biot number.  In both panels, the dimensionless laser radius is $w = w^* / h^* = 2$. }
    \label{fig:Angle}
\end{figure}

When a beam is irradiated by a laser, differential
thermal expansion along the thickness of the beam leads to a localised
bending moment.  When the ends of the beam are free, the bending moment
plays the role of an active hinge, causing the beam to bend into the
shape of a V~\cite{Light,Cao2022}.  
A key quantity of interest is the fold angle $\phi$, which measures the
angle of elevation of one arm of the V relative to the other.  Fold angles of $\phi = 0$, $\pi/2$, and $\pi$ correspond, respectively, to the undeformed beam, the two arms of the V being perpendicular, and the arms of the V folding into each other.  The fold angle can be calculated from the
jump condition \eqref{red:bc_first} and is given by
\begin{equation}
    \phi = \arctan\LR{\Gamma_1 M}, 
\end{equation}
which can be approximated as $\phi \approx \Gamma_1 M$ for small
angles.  
Dimensionally, the quantity $\Gamma_1$ can be written as
\begin{equation}
    \Gamma_1 = \frac{\gamma_1^*\eta_{\text{th}}\beta^*h^*L^*P^*}{\pi (w^*)^2 k^*},
    \label{eqn:dim_Gamma_1}
\end{equation}
whereas the dimensionless thermal moment depends non-trivially on
$w^* / h^*$, $\beta = \beta^* h^*$, and $\Bi = H^* h^* / k^*$.
In the small-angle regime, increasing the laser power, $P^*$, will lead
to a linear increase in the fold angle.  However, the dependence of the fold angle on the attenuation coefficient, $\beta$, 
which can be controlled by adding photothermal agents to the material, is more subtle. 
When optical attenuation is weak, $\beta \ll 1$, the leading-order
inner thermal problem can be expanded in powers of $\beta$.  The 
leading-order contribution (in terms of $\beta$) to the inner temperature field will
be symmetric about $X_2 = 0$ and hence not produce a thermal moment.
The $O(\beta)$ contribution to the temperature will be asymmetric due to
the gradient in light intensity entering at this order
and will produce a thermal moment of size
$O(\beta)$.  
The heat that is generated due to light absorption is $O(\beta)$ as
well.  Thus, when $\beta \ll 1$, the small thermal moment acts 
cooperatively with the small heat generation to cause the fold angle
to increase quadratically with the attenuation coefficient, 
$\phi = O(\beta^2)$, as confirmed numerically; see Fig.~\ref{fig:Angle}~(a).
As the attenuation coefficient increases, the quadratic growth
saturates and $\phi \to \pi$, corresponding to a completely folded
beam.  For $\beta \gg 1$, the light intensity decays to zero over
a thin region of width $O(1/\beta)$ located at the upper surface of the 
beam.  The localised intensity still induces a vertical temperature
gradient across the height of the beam, although this gradient is small
and $O(1/\beta)$ in size.  The resulting thermal moment is then
$O(1/\beta)$; see Fig.~\ref{fig:Angle}~(b).  Thus, as $\beta$ increases,
the linear increase in heat 
generation is perfectly counteracted by a 
decrease in thermal moment to produce a constant fold angle.  
In the case of perfect heat exchange with the environment,
corresponding to $\Bi \to \infty$, the rapid conduction of heat leads
to a thermal moment that scales as $M = O(1/\beta^3)$ as $\beta \to \infty$.  In this case, the fold angle is expected to decay to zero for
large $\beta$. 

\subsection{Snap-through triggered by laser irradiation}

We now use the model to investigate the light-driven snap-through of a pre-buckled hydrogel beam as reported in \cite{Light}.  Specifically, 
we study how the equilibrium configurations of the beam 
change as the dimensionless
longitudinal thermal strain, $\Gamma_1$, varies.  In experimental terms,
varying $\Gamma_1$ is possible by changing the power of the 
laser; see \eqref{eqn:dim_Gamma_1}.

The conditions for which Euler bucking occurs are first obtained by
solving \eqref{eq: beam model} with the thermal loading terms set to zero, $\Theta=M=0$.  A non-trivial solution for the transverse displacement only exists when $\lambda>4\pi^2$, in which case the displacement is given by
\begin{equation}
    v_E(X_1) = -\frac{\sqrt{\lambda-4\pi^2}}{\pi}\left[1+\cos(2\pi X_1)\right],
\end{equation}
where the negative sign reflects the downwards-pointing direction of the $X_2$ coordinate. 
When a thermal load is applied, the solution for the transverse displacement
becomes
\begin{align}
    \begin{split}
        v_{\pm} = v = -\frac{\Gamma_1M}{2\mu}\bigg[&\frac{\cos(\mu X_0)}{\sin\LR{\frac\mu2}}\LR{1-\cos\LR{\frac{\mu}{2}}\cos(\mu X_1)-\sin\LR{\mu|X_1-X_0|}}\\&+\frac{\sin(\mu X_0)}{\mu\cos\LR{\frac{\mu}{2}}-2\sin\LR{\frac{\mu}{2}}}\LRBR{\LR{\mu\sin\LR{\frac{\mu}{2}}+2\cos\LR{\frac{\mu}{2}}}\sin(\mu X_1)-2\mu X_1}\bigg],
    \end{split}
\end{align}
which is valid in both outer regions. The constant $\mu$ is determined from the longitudinal 
force balance, which can be written as
\begin{equation}
    F(\mu,\Gamma_1):=\mu^2-\lambda-\Gamma_1\Theta+\frac12
        \int_{-1/2}^{1/2}\left(\odv{v}{X_1}\right)^2\,\mathrm{d}X_1 = 0
\end{equation}
for fixed end shortening $\lambda$. The transverse displacement at the mid-point of
the bean is given by
\begin{equation}\label{midpoint disp}
    v(0) = -\frac{\Gamma_1M}{2\mu}\LR{\tan\LR{\frac\mu4}\cos(\mu X_0)-\sin(\mu|X_0|)},
\end{equation}
which we use to characterise how the equilibria vary with
$\Gamma_1$.  To determine the critical longitudinal stress $\mu$ and thermal strain
$\Gamma_1$ leading 
to snap through, we use the limit-point condition \cite{thompson1973} to find
\begin{equation}
    F(\mu,\Gamma_1)=0\wand \pdv{F}{\mu}(\mu,\Gamma_1)=0.
    \label{eqn:snap_through}
\end{equation}

\begin{figure}
    \centering
    \includegraphics[width=0.95\linewidth]{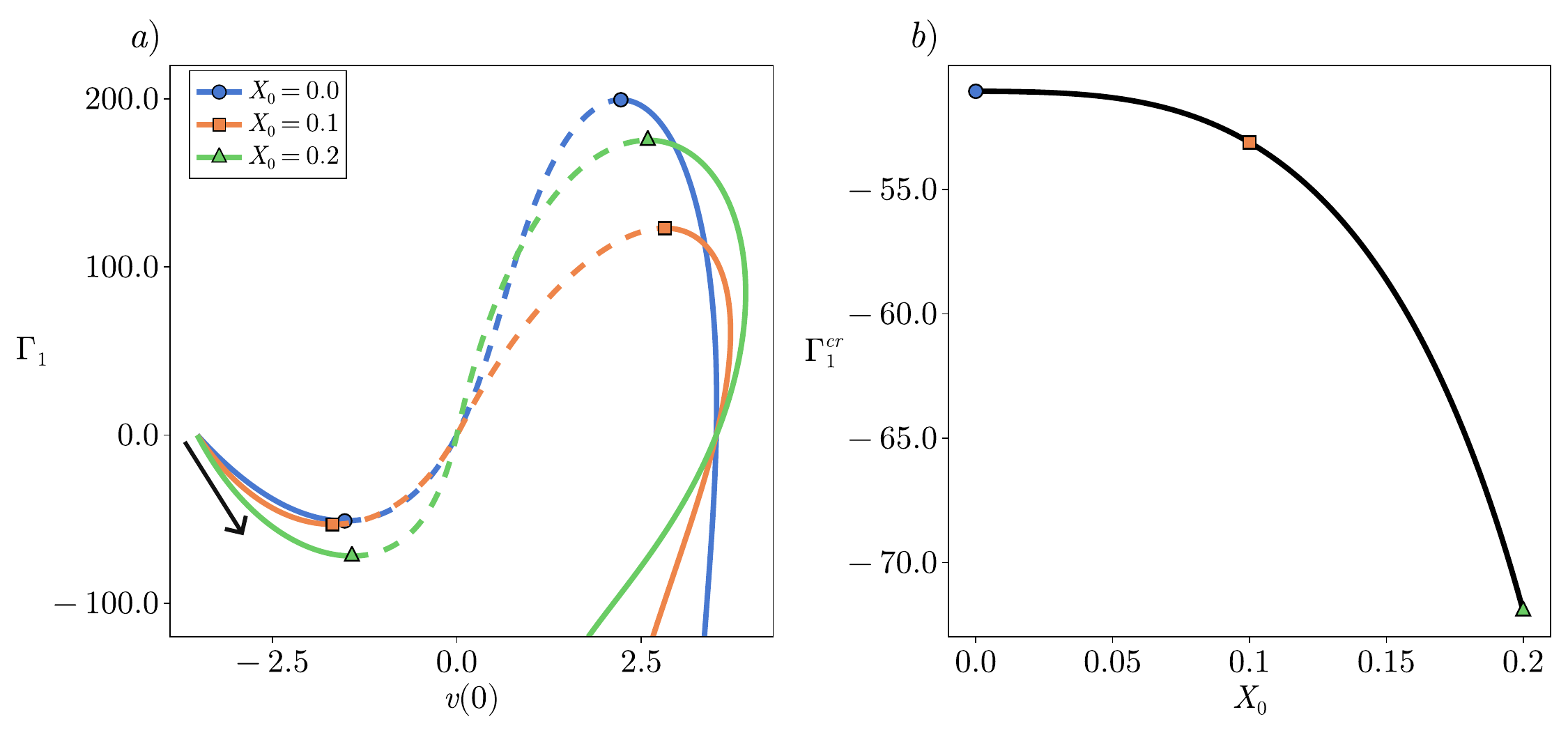}
    \caption{Laser-induced snap-through of a thermoelastic beam.
    (a) Bifurcation diagrams showing the how the 
    transverse displacement
    of the mid-point of the beam $v(0)$ changes with
    $\Gamma_1$.  For this parameter set, increasing the
    laser power decreases $\Gamma_1$, as shown by the
    black arrow.  Solid and dashed lines represent linearly
    stable and unstable equilibrium configurations, 
    respectively.  (b) The critical value of $\Gamma_1$
    at which snap through occurs as a function of the
    point of irradiation $X_0$.  In both plots, the
    non-dimensional parameters are $\lambda = 70$, 
    $w = 2$, $\beta = 10$, and $\Bi = 10$.}
    \label{fig:Snap-through}
\end{figure}

To study photo-induced snap-through, we fix $\lambda = 70$
so that with no thermal loading ($\Gamma_1 = 0$) the beam
is in a buckled configuration with $v(0) \simeq -3.5$.  
If the laser power increases, corresponding to a decrease in
$\Gamma_1$ due to the negative coefficient of thermal expansion
for this parameter set, then $v(0)$ increases and the mid-point
of the beam begins to move closer to its original position.  As $\Gamma_1$ decreases
further, the curve of equilibria folds back on itself at a
critical value $\Gamma_{1}^{cr}$, at which point a 
limit-point bifurcation occurs and the branch of equilibria
becomes linearly unstable;
see Fig.~\ref{fig:Snap-through}~(a).
Decreasing $\Gamma_1$ beyond $\Gamma_{1}^{cr}$ induces snap through and the beam rapidly evolves towards 
a new and remote equilibrium with a positive value of $v(0)$,
corresponding to an everted configuration.  

The snap-through point depends strongly on the longitudinal
point of irradiation, $X_0$.  By computing bifurcation diagrams
with $X_0 = 0$, $0.1$, and $0.2$, we find that $\Gamma_{1}^{cr}$
decreases with $X_0$ (Fig.~\ref{fig:Snap-through}~(a)).
Numerically solving \eqref{eqn:snap_through} across a range of 
$X_0$ shows that $\Gamma_{1}^{cr}$ monotonically decreases with
$X_0$ (Fig.~\ref{fig:Snap-through}~(b)).  Thus, offsetting the laser from the mid-point of
the beam provides a means to stabilise the system, as 
triggering snap through requires a greater laser power $P^*$.


\section{Conclusion}\label{Conclusion}

In this paper, we derive a reduced model for a photoresponsive thermoelastic beam subject to localised heating.  Motivated by recent studies of contactless actuation of hydrogel beams,
we assume that heating is due to a laser passing through the
material.  By leveraging the small aspect ratio of the beam alongside the localisation
of the heating, a matched asymptotics problem is formulated in which the
domain of the problem is decomposed into an inner thermoelastic heated region and
two isothermal outer regions governed by a nonlinear beam
equation with von K\'arm\'an strains.
By resolving the inner region and using
asymptotic matching, asymptotically consistent jump
conditions across the inner region are obtained for the beam equations.  The jump conditions reveal that the heating, despite being localised, still
has a leading-order impact on the length of the beam, a feature that has not been
reported before.  Furthermore, the analysis illustrates how thermal moments,
generated by asymmetric temperature profiles, induce local rotation of material
elements and bending moments, as captured through non-trivial jump conditions on 
the first and third derivatives of the transverse displacements.  

The reduced model is used to study two canonical problems involved photo-thermal
responsive hydrogels.  By considering the V-shaped deformation caused by local
heating of free hydrogel beams, we show that the fold angle increases linearly
with laser power and quadratically with the optical attenuation coefficient.  
Thus, increasing the absorption of light through additives can provide a facile means
of increasing the actuation range of material.  We then examine how light-induced 
snap through can be induced in pre-buckled hydrogels with clamped ends.  The critical conditions are sensitive to the
point of irradiation.  When the laser is offset from the
mid-point of beam, $X_1^* = 0$, a greater laser power is
required to induce snap through.  In both problems, only the equilibrium response of the beam is considered.  Future studies
could focus on the dynamics of thermally driven shape change.  For example, the model developed here can be used to explore the interplay between the elastic and thermal time scales, which is expected to control the rate of snap through~\cite{Radisson2023,Simpkins2026}.




The analysis is developed under the assumption that
the Biot number is $O(1)$.  In this case, the cooling that occurs
in the inner region is sufficient to prevent heat from propagating
into the outer regions.  When $\Bi \ll 1$, there is weak heat exchange with the environment.  Therefore, the heat generated in the inner region is transferred to the outer region; the localised
heating induces a global temperature increase.  The asymptotic 
reduction of the model when $\Bi \ll 1$ is, consequently, more
complex and will be the topic of a future publication.

A second key assumption in the analysis is that the laser strikes the
transverse mid-point of the upper surface ($X_3^* = 0$).  Due to the symmetry
of the temperature profile and lack of a thermal moment in the
$X_3$ direction, the beam will undergo
a plane-strain deformation, allowing the model to be further reduced.  Offsetting the
irradiation point would lead to a much richer space of possible deformations
that could be captured by extending the asymptotic framework developed here.

This work provides a systematic methodology for deriving reduced models of slender structures undergoing localised stimulation.  
Although we focus on photothermal actuation, the governing equations
can be used to study the response of a thermoelastic beam to any localised heat 
source.  Moreover, similar asymptotic analyses can be used to capture the response
of materials to other types of localised changes in volume, 
including hydrogel swelling/deswelling due to chemical, electrical, and magnetic stimuli.  By providing
a framework to reduce the computational cost of simulating complex, responsive
materials, new advances in field such as soft robotics and 4D printing can be
made.

\section*{Acknowledgements}

MT is a member of the Gruppo Nazionale di Fisica Matematica (GNFM) of the Istituto Nazionale di Alta Matematica (INdAM).  
MGH was partially supported by the Engineering and Physical Sciences Research Council (Grant No. UKRI093).

\appendix

\section{The longitudinal force balance}\label{app:axial}

When clamped boundary conditions are imposed, it is convenient
to determine the mean longitudinal stress $\B{P_{11}^{(0)}}$ in
terms of the imposed end shortening $\lambda$.  In the outer
regions, the mean longitudinal stress is given by
\eqref{P11 outer expression}.  Integrating the mean longitudinal stress over the
outer regions, summing the results, and using 
$\B{P_{11,+}^{(0)}} = \B{P_{11,-}^{(0)}} = \B{P_{11}^{(0)}}$
leads to
\begin{equation}
    \B{P_{11}^{(0)}} = \int_{-\frac12}^{X_0}\left(\pdv{U_-}{X_1}+\frac12\pdv{u_{\beta,-}^{(0)}}{X_1}\pdv{u_{\beta,-}^{(0)}}{X_1}\right)\text{d}X_1+\int_{X_0}^{\frac12}\left(\pdv{U_+}{X_1}+\frac12\pdv{u_{\beta,+}^{(0)}}{X_1}\pdv{u_{\beta,+}^{(0)}}{X_1}\right)\text{d}X_1,
    \label{axial:P11}
\end{equation}
From the boundary and jump conditions for $U_\pm$ given
by \eqref{U jump} and \eqref{outer_bc_clamped}, we have
\begin{equation}
    \int_{-\frac12}^{X_0}\pdv{U_-}{X_1}\,\text{d}X_1+\int_{X_0}^{\frac12}\pdv{U_+}{X_1}\,\text{d}X_1 = -\lambda-\Gamma_1\Theta.
    \label{axial:length}
\end{equation}
Substitution of \eqref{axial:length} into \eqref{axial:P11}
leads to \eqref{outer:P11_soln}.

\section{Solutions to the inner thermal problem}
\label{app:thermal}

A semi-analytical solution for the inner thermal problem can be
obtained using eigenfunction expansions. 
Let $\varphi(X_2, X_3)$ solve the transverse eigenvalue problem 
\begin{equation}
    \frac{\p^2\varphi}{\p X_\alpha \p X_\alpha} = -\kappa^2\varphi,
\end{equation}
with boundary conditions
\begin{equation}
    \pdv{\varphi}{X_\alpha}\pm\Bi\, \varphi=0 \wat X_\alpha =\pm\frac12,
\end{equation}
where $\Bi > 0$.
Because the cross-section is square and the same Newton cooling condition is imposed on each transverse face, the eigenfunctions can be expressed as products of one-dimensional eigenfunctions. We let $\psi_j(X)$ solve
\begin{equation}
    \odv[2]{\psi_j}{X}+\nu_j^2\psi_j=0,
\end{equation}
with boundary conditions 
\begin{equation}
    \odv{\psi_j}{X}\pm\Bi\,\psi_j=0 \wat X = \pm \frac{1}{2}.
\end{equation}
The modes are chosen to be orthonormal on so that
\begin{equation}
    \int_{-1/2}^{1/2}\psi_n(X)\psi_m(X)\text{d}X=\delta_{nm}.
\end{equation}
The corresponding two-dimensional eigenfunctions are 
\begin{equation}
    \varphi_{nm}(X_2, X_3)=\psi_n(X_2)\psi_m(X_3),
\end{equation}
with eigenvalues 
\begin{equation}
    \kappa_{nm}^2=\nu_n^2+\nu_m^2.
\end{equation}
The one-dimensional eigenfunctions can be ordered by parity. We choose to index so that the even-numbered modes are even functions of $X$, while the odd-numbered modes are odd functions. Therefore we have 
\begin{equation}
    \psi_{2q}(X)=C_{2q}\cos(\nu_{2q}X) \wand \psi_{2q+1}(X)=C_{2q+1}\sin(\nu_{2q+1}X),
\end{equation}
where $q=0,1,\ldots$, with eigenvalues that are implicitly defined by 
the equations
\begin{equation}
    \nu_{2q}\tan\LR{\frac{\nu_{2q}}{2}}=\Bi\wand \nu_{2q+1}\cot\LR{\frac{\nu_{2q+1}}{2}}=-\Bi.
    \label{app:eig}
\end{equation}
In practice, the eigenvalues are obtained by solving \eqref{app:eig} 
numerically.
The coefficients of the eigenfunctions are given by
\begin{equation}
    C_{2q}=\LR{\frac12+\frac{\sin(\nu_{2q})}{2\nu_{2q}}}^{-\frac12} \wand C_{2q+1}=\LR{\frac12-\frac{\sin(\nu_{2q+1})}{2\nu_{2q+1}}}^{-\frac12}.
\end{equation}
We then write both the temperature distribution and the light intensity as an eigenfunction expansion,
\begin{subequations}
\begin{align}
    \theta_{\i}^{(0)}(\XM,X_2, X_3,t)&=\sum_{n,m} \theta_{nm}(\XM,t)\varphi_{nm}(X_2, X_3),
    \\
    I_{\i}^{(0)}(\XM,X_2, X_3) &= \sum_{n,m} I_{nm}(\XM)\varphi_{nm}(X_2, X_3),
\end{align}
\end{subequations}
where the modal amplitudes for the intensity are given by
\begin{equation}
    I_{nm}(\XM)=\int_{-1/2}^{1/2} \int_{-1/2}^{1/2} I_{\i}^{(0)}(\XM,X_2, X_3)\varphi_{nm}(X_2, X_3)\,\mathrm{d}X_2\mathrm{d}X_3
\end{equation}
and the thermal amplitude $\theta_{nm}$ are obtained from
\begin{align}
\pdv[1]{\theta_{nm}}{t} = \pdv[2]{\theta_{nm}}{\XM} - \kappa^2_{nm} \theta_{nm} + I_{nm}.
\label{eqn:thermal_modal}
\end{align}
Since the mechanical problem only requires integrated thermal
quantities, we define
\begin{subequations}
\begin{align}
    \Theta_{nm}(t) = \int_{-\infty}^\infty \theta_{nm}(\XM,t)\text{d}\XM
    \wand
    \mathcal{I}_{nm}=\int_{-\infty}^\infty I_{nm}(\XM)\text{d}\XM.
\end{align}
\end{subequations}
By integrating \eqref{eqn:thermal_modal} over the inner region
and imposing $\p \theta_{nm} / \p \XM \to 0$ as $\XM \to \pm \infty$ by matching, we find that
\begin{equation}
    \odv{\Theta_{nm}}{t}+\kappa^2_{nm}\Theta_{nm}=\mathcal{I}_{nm},
\end{equation}
which has initial condition $\Theta_{nm}(0)=0$. 
Since $\Bi > 0$, $\kappa_{nm}^2 > 0$ and hence
\begin{equation}
    \Theta_{nm}(t)=\frac{\mathcal{I}_{nm}}{\kappa_{nm}^2}\LR{1-e^{-\kappa^2_{nm}t}}.
    \label{eqn:Theta_nm}
\end{equation}
The mean temperature in the inner region and the thermal moment
can then be calculated as
\begin{subequations}
\begin{align}
\Theta(t) &= \sum_{nm} \Theta_{nm}(t) \B{\varphi_{nm}},
\\
M_2(t) &= \sum_{nm} \Theta_{nm}(t) \B{X_2 \varphi_{nm}},
\end{align}
\end{subequations}
where the overbar denotes the cross-section average defined
in \eqref{eqn:average}.   As $t \to \infty$, the mean
temperature and thermal moment approach a steady state
given by
\begin{subequations}\label{steady_state}
\begin{align}
\Theta^{\infty} &= \sum_{nm} \Theta_{nm}^{\infty} \B{\varphi_{nm}},
\\
M_2^{\infty} &= \sum_{nm} \Theta_{nm}^{\infty} \B{X_2 \varphi_{nm}},
\end{align}
\end{subequations}
where $\Theta_{nm}^{\infty} = \mathcal{I}_{nm} / \kappa_{nm}^2$
from \eqref{eqn:Theta_nm}.
If $\Bi=0$, then there is a zero eigenvalue, $\kappa_{00}=0$.
The mean of the corresponding thermal mode grows according to
\begin{equation}
    \Theta_{00}(t) = \mathcal{I}_{00}t.
\end{equation}
This reflects the fact that, without heat loss, the total heat content grows linearly in time.

\section{Parameter values}\label{sec:Parameter values}

The parameters are based on light-responsive PNIPAM
hydrogels developed by Dai et al.~\cite{Light}.  PNIPAM
hydrogels deswell and lose volume when heated, which we
capture using a negative coefficient of thermal expansion in the longitudinal
directions.
Table~\ref{tab:dim variables} provides the parameter values used
throughout the analysis.  

\begin{table}[]
    \centering
        \caption{A list of dimensional parameter values used throughout
        the analysis obtained from Refs~\cite{Light, Brunner2024}.}
        \vspace{0.25em}
        \begin{tabular}{||c|c|c||}
        \hline
            Parameter  & Value & Units \\
        \hline
            $L^*$  & $0.02$ & m \\
            $h^*$  & $0.002$ & m \\
            $\theta_0^*$  & $300$ & K \\
            $k^*$  & $0.6$ & W m$^{-1}$ K$^{-1}$ \\
            $E^*$  & $6.2\times 10^5$ & Pa  \\
            $c_\theta^*$ &  $4.2\times 10^3$ & J kg$^{-1}$ K$^{-1}$ \\
            $\gamma_1^*$  & $-1.7\times 10^{-2}$ & K$^{-1}$  \\
            $P^* / [\pi (w^*)^2]$ & $3\times10^4$ & W m$^{-2}$ \\
            $\rho_0^*$ & $10^3$ & kg m$^{-3}$  \\
            $\beta^*$ & $10^2$ & m$^{-1}$ \\
        \hline
        \end{tabular}
        \label{tab:dim variables}
\end{table}

Using these parameters, the non-dimensional numbers appearing in the 
model can be quantified or estimated.  The aspect ratio of the
beam is $\delta = h^* / L^* = 0.1$.  The expected temperature increase
is $\Delta \theta^* = 20$~K, resulting in a small relative temperature
increase $\varepsilon = \Delta \theta^* / \theta_0^* \simeq 0.067$.
The dimensionless Young's modulus is also found to 
be small, $\mathcal{E} \simeq 0.09$, which provides further justification
for neglecting the mechanical contributions to the entropy, as defined by \eqref{entropy}.  
The dimensionless thermal strain is $-\gamma_1 \Delta \theta^* \simeq  -0.34$ and is similar in size to $\delta$, justifying the distinuished
limit considered in Sec.~\ref{Asymptotics}.  The ratio between the
elastic and thermal diffusion time scales is 
$\tau \simeq 3 \times 10^{-6}$, resulting in a value of
$\epsilon = 3 \times 10^{-4}$.  The dimensionless attenuation coefficient
is $\beta = \beta^* h^* \simeq 0.2$.  The smallness of $\beta$ indicates
that the gradient in light intensity will be weak.
The heat transfer coefficient $H^*$ was not measured; however, if we
assume that $H^*$ is between 10 to 1000 W m$^{-2}$ K$^{-1}$, then the
Biot number will range from 0.03 to 0.3. 



\bibliographystyle{ieeetr}
\bibliography{sample}

\end{document}